\documentclass[useAMS,usenatbib]{mn2e}

\usepackage{graphicx}
\usepackage{amsmath}
\usepackage{amssymb}
\usepackage{url}
\usepackage{subfigure}
\usepackage{natbib}

\makeatletter{}

\def\fp{\hbox{$.\!\!^{\reset@font\scriptscriptstyle\r@mn{p}}$}}

\long\def\crap#1{}

\def\<{\langle }
\def\>{\rangle }

\title[Simulation pipeline for HARMONI on the E-ELT]{HSIM: a simulation pipeline for the HARMONI integral field spectrograph on the European ELT}
\author[Zieleniewski et al.]{S. Zieleniewski$^{1}$\thanks{E-mail:
simon.zieleniewski@physics.ox.ac.uk} N. Thatte$^{1}$ S. Kendrew$^{1}$ R. C. W. Houghton$^{1}$\newauthor
A. M. Swinbank$^{2}$ M. Tecza$^{1}$ F. Clarke$^{1}$ T. Fusco$^{3,4}$\\
$^{1}$Astrophysics, Denys Wilkinson Building, Keble Road, Oxford, OX1 4RH\\
$^{2}$Institute for Computational Cosmology, University of Durham, South Road, Durham, DH1 4LH\\
$^{3}$ONERA, BP 72, F-92322, Chatillon, France\\
$^{4}$Aix Marseille University, CNRS LAM (Laboratoire d'Astrophysique de Marseille) UMR 7326, 13388 Marseille, France}
\begin{document}

\date{Last updated 18/08/15}

\pagerange{\pageref{firstpage}--\pageref{lastpage}} \pubyear{2014}

\maketitle

\label{firstpage}

\begin{abstract}
We present {\sc hsim}: a dedicated pipeline for simulating observations with the HARMONI integral field spectrograph on the European Extremely Large Telescope. {\sc hsim} takes high spectral and spatial resolution input data-cubes, encoding physical descriptions of astrophysical sources, and generates mock observed data-cubes. The simulations incorporate detailed models of the sky, telescope and instrument to produce realistic mock data. Further, we employ a new method of incorporating the strongly wavelength dependent adaptive optics point spread functions. {\sc hsim} provides a step beyond traditional exposure time calculators and allows us to both predict the feasibility of a given observing programme with HARMONI, as well as perform instrument design trade-offs. In this paper we concentrate on quantitative measures of the feasibility of planned observations. We give a detailed description of {\sc hsim} and present two studies: estimates of point source sensitivities along with simulations of star-forming emission-line galaxies at $z\sim$\,2--3. We show that HARMONI will provide exquisite resolved spectroscopy of these objects on sub-kpc scales, probing and deriving properties of individual star forming regions.
\end{abstract}

\begin{keywords}
instrumentation: spectrographs - instrumentation: adaptive optics - galaxies: high-redshift - galaxies: kinematics and dynamics
\end{keywords}

\section[introduction]{Introduction}

As the current suite of 8 m class telescopes are being pushed to their limits, it is becoming more important to prepare for and define the scientific programmes to be undertaken by the next generation of extremely large telescopes. The European Extremely Large Telescope (E-ELT) is a 39 m diameter telescope being constructed by the European Southern Observatory (ESO) for operations commencing within the next decade. Upon completion, this facility will provide the angular resolution and light gathering power to revolutionise our current understanding in many areas of astrophysics. However, as telescope and instrumentation projects enter such ambitious scales of size and complexity of design, it is becoming imperative to accurately quantify performance prior to construction.

Consequently, several recent integral field spectrograph (IFS) instrumentation projects have developed detailed instrument simulation models, including KMOS \citep{Lorente2008}, MUSE \citep{Jarno2008}, JWST/NIRSpec \citep{Piqueras2010, Dorner2011}, and EAGLE/ELT-MOS \citep{Puech2008, Puech2010a, Puech2010b}. The most rigorous simulation method involves using the full optical design of the instrument to propagate photons to the detector plane, before using data reduction software to reduce the simulations akin to real data. While this full end-to-end technique is possible for currently operational instruments and those at an advanced stage of design, it is a level above what is required at an early stage of development. 

As part of the concept design for an ELT multi-object spectrograph (MOS), \citet{Puech2010b} developed an IFS instrument simulator that processed input data-cubes, generating mock observation cubes assuming a perfect data reduction process. The simulator encoded all the sky, telescope and adaptive optics (AO; through the point spread function) and instrument parameters, returning mock observations containing source and background flux as well as noise contributions. They were able to analyse the feasibility of a key MOS program to observe star forming galaxies at $z\sim6$.

The High Angular Resolution Monolithic Optical and Near-infrared Integral field spectrograph (HARMONI) is a proposed instrument selected as one of two first-light instruments by ESO \citep{Thatte2010}. HARMONI is being designed as a single-field, visible and near infra-red (NIR) IFS, and will provide a range of spatial pixel (spaxel) scales and spectral resolving powers, which permit the user to optimally configure the instrument for a wide range of science programs; from ultra-sensitive to diffraction limited, spatially resolved, physical (morphology), chemical (abundances and line ratios) and kinematic (line-of-sight velocities) studies of astrophysical sources. HARMONI will be compatible with two modes of adaptive optics (AO) allowing it to tackle a broad range of problems across astrophysics, including: a) the physics of mass assembly of galaxies at high redshifts, b) resolved studies of stellar populations in distant galaxies, and c) detecting and weighing intermediate mass black holes in nearby galaxies or globular clusters. The full design specifications are listed in table 1. The instrument will commence its preliminary design phase in 2015 and so here we undertake detailed simulations of the performance of HARMONI.

\begin{table*}
 \centering
 \begin{minipage}{115mm}
  \caption{Current design parameters for HARMONI.}
  \begin{tabular}{@{}lr}
 \hline
 \vspace{3pt}
Wavelength range & 0.47 - 2.45 $\mu \text{m}$ \\

Spatial scales (FoV) & 4 mas ($0.86'' \times0.61''$), 10 mas ($2.14'' \times1.52''$)\\
\vspace{5pt}
 & 20 mas ($4.28'' \times3.04''$), $60\times30$ mas ($9.12'' \times6.42''$)      \\

Spectral Resolution (\& waveband) & 3500 ($V+R$; $Iz+J$; $H+K$)\\
 & 7500 ($Iz$; $J$; $H$; $K$) \\
 \vspace{5pt}
  & 20000 (half of each NIR band)    \\
 \vspace{3pt}
Temperature & 120 K cryostat, 40 K detectors     \\

Detectors & Teledyne HgCdTe 4k $\times$ 4k for NIR\\
 \vspace{5pt}
 & 4k $\times$ 4k CCD for visible\\
 \vspace{3pt}
Throughput & Target $\geq 30\%$ average\\

AO modes & SCAO, LTAO, seeing-limited     \\
\hline
\end{tabular}
\end{minipage}
\end{table*}

In this paper we present our development of the instrument simulation pipeline {\sc hsim} for the HARMONI instrument that will be used to quantify the performance. {\sc hsim} improves on the method developed by Puech et al. in several key areas, including spectral line spread convolutions, and a wavelength dependent telescope (plus adaptive optics) point spread function model. We detail the components of {\sc hsim} and show the importance of the key stages and their effects on observations. We then showcase the pipeline with two studies: calculations of point source sensitivities for the available observing modes, and a case study of measuring kinematics of star forming galaxies at $z\sim$\,2--3.

The paper is organised as follows: Sections~\ref{motivation} and~\ref{methodology} detail the motivation, goals and methodology of the simulations; in Section~\ref{overview} we present an overview of the pipeline stages; in Section~\ref{psfs} we present our point spread function parameterisation and demonstrate its importance; Section~\ref{verification} shows our verification crosschecks against existing software; in Section~\ref{sensitivities} we present predicted point source sensitivities; Section~\ref{swinbank} presents our high redshift galaxy simulations; and we finally conclude in Section~\ref{conclusion}.

Throughout this paper we adopt a flat $\rmn{\Lambda}$CDM cosmology with $H_0 = 70\,\rmn{km}\,\rmn{s}^{-1}\,\rmn{Mpc}^{-1}$, $\Omega_m = 0.3$ and $\Omega_{\Lambda} = 0.7$. We use AB magnitudes throughout unless otherwise stated.

\section{Motivation and Goals}
\label{motivation}

Developing an instrument simulation pipeline serves several important purposes:
\\
a) it gives a quantitative understanding of the feasibility of observing programs, characterising both the capabilities and limitations of an instrument,\\
b) it allows for performance trade-offs between differing instrument designs or configurations.\\

The first point is important since it will allow observing time on the E-ELT to be more efficiently utilised. Quantifying performance usually consists of determining the exposure time required to achieve a particular signal-to-noise ratio (SNR) for a given target. However, an instrument simulation pipeline allows us to go one step beyond exposure time calculators and ascertain the precision with which we can derive a number of key physical parameters for particular science cases. The input data-cubes can be built to encode as many physical characteristics as desired. The mock output data represents that of real observations and so it is possible to use identical analysis methods as used on real data (or permits users to develop and verify analysis methods prior to taking the real data). This allows the user to specifically determine the feasibility of their science goal (e.g. measuring kinematics, black hole masses, stellar chemical abundances). It also provides a direct comparison between the scientific capabilities of current generation instrumentation and those of the E-ELT era.

Designing a self-contained and modular pipeline means that all simulations are kept consistent within the current design of the instrument. The pipeline can easily be modified to reflect any changes in the instrument design (e.g. changing spaxel scales, spectral resolving power etc), thus allowing exploration of the second point above. It will also be possible to explore instrument effects such as imperfect sky subtraction and spectrograph throughput variations, and give quantitive constraints on such calibration issues. In this paper we concentrate on the accuracy with which physical parameters can be measured for a specific science case.

\section[simulation methodology]{Simulation Methodology}
\label{methodology}

The procedure for performing simulations can be broken down into three parts:\\
a) An input data-cube encoding the physical characteristics of the object. This can include properties such as kinematics, absorption/emission lines, dynamics, morphology, chemical abundances etc;\\
b) A simulation pipeline that takes the input and adds all the first-order sky, telescope, instrument, and detector effects, as well as random and systematic noise, creating an output mock-observed data-cube. The user is able to choose suitable observing parameters to be able to explore the instrument and telescope modes;\\
c) The analysis of the output mock-observed data to enable an understanding of what information can be extracted from the observation, i.e. how well can the properties of the input data-cube be recovered in the output cube.
\\
\\
Within this framework there are two areas of parameter space to be explored: the physical characteristics of the objects (e.g. morphology, mass, redshift, magnitude, line strengths) as well as the observational characteristics of the telescope, and instrument (e.g. spatial scale, exposure time, resolving power, emissivity, AO correction). Consequently the combined parameter space is vast, and a full exploration is simply not possible. For the purposes of this paper, we have firstly pursued point source sensitivity calculations for all current instrument configurations, which gives a general overview of the performance of the instrument design. We then present simulations of a representative case of $z\sim$\,2--3 emission-line galaxies.

\section[simulation pipeline overview]{Simulation Pipeline Overview}
\label{overview}

\begin{figure}
\centering
\resizebox{1.04\columnwidth}{!}{\includegraphics{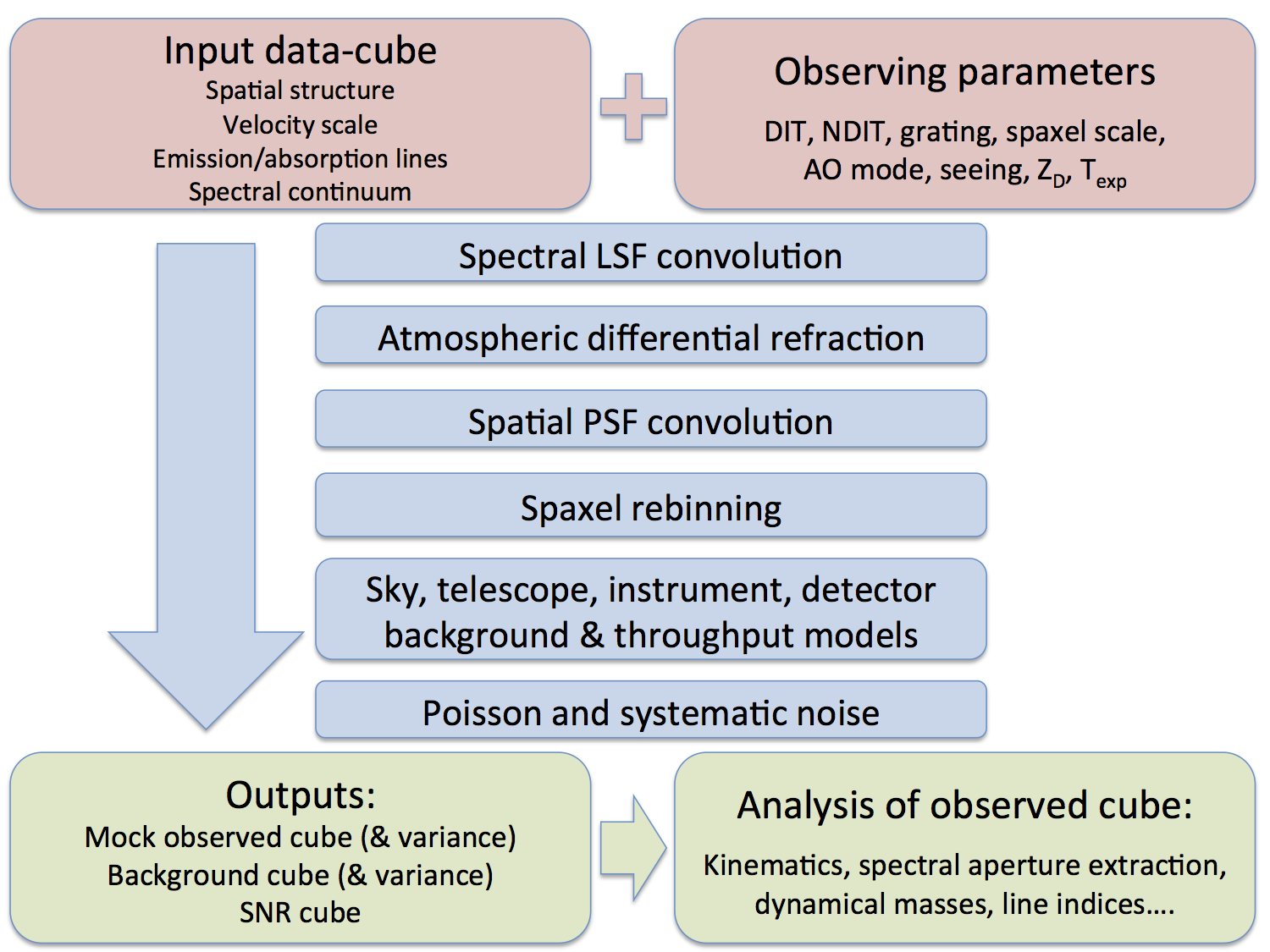}}
\caption{Flowchart of the simulation process. Input data-cubes provide all the physical details of the object. The cube is then `observed' for given instrument, telescope and site parameters. The observing process adds all first-order telescope and instrument effects, as well as random and systematic noise. The output cube represents a perfectly reduced data-cube, and can be analysed exactly like real data.}
\label{flowchart}
\end{figure}

We first presented the simulation pipeline in \citet{Zieleniewski2014} and here we provide an overview of each stage, highlighting several novel features. The overall process is presented in Fig. \ref{flowchart}. {\sc hsim} is written in {\sc python} and makes use of the {\sc astropy} package \citep{Astropy2013}, as well as the {\sc numpy} and {\sc scipy} packages. It handles input and output data in {\sc fits} format, as this is the usual file format for 3D spectrographs. The input data-cube is uploaded as a 3D array of two spatial and one spectral dimension. The {\sc fits} headers contain all the relevant information about the object (e.g. spatial and spectral sampling, flux units, spectral resolution). The user chooses suitable observing parameters and the data-cube is then processed through several steps:

\subsection{Line spread function convolution}

The spectral resolution of the input cube is degraded to the output resolution depending on the chosen grating. This is achieved by convolving the spectral dimension with a Gaussian line-spread function (LSF). We use a Gaussian of constant full width half maximum (FHWM) as a reasonable approximation for the resolution of a slit-width limited grating spectrometer. We ensure the output resolution of the cube is consistent by generating a LSF of width $\Delta\lambda_{\text{conv}}$ given by
\begin{equation} \Delta\lambda_{\text{conv}} = \sqrt{(\Delta\lambda_{\text{out}})^2 - (\Delta\lambda_{\text{in}})^2},\end{equation}
where $\Delta\lambda_{\text{in}}$ is the spectral resolution of the input cube specified in the {\sc fits} header (by the user), and $\Delta\lambda_{\text{out}}$ is the output resolution given by the chosen grating. The spectral dimension of the cube is then sampled (with a minimum of two pixels per FWHM to approximate the Nyquist limit). If the input spectral resolution is coarser than the grating resolution ($\Delta\lambda_{\text{in}} > \Delta\lambda_{\text{out}}$) then the user can choose to either convolve with a LSF of width $\Delta\lambda_{\text{out}}$ or perform no spectral convolution. The former option is useful in situations where the input data-cube contains a spectrum where the resolution is limited by e.g. stellar velocity dispersion in galaxies, and the spectrum still needs to incorporate the instrumental effect. The option to ignore the spectral convolution is always available if it is not required (e.g. if the input data-cube has been pre-formatted spectrally).

\subsection{Atmospheric differential refraction}

The effect of atmospheric differential refraction (ADR) is added according to the equations of \citet{SchubertWalterscheild2000} and \citet{Roe2002}. The angle in radians between the true zenith distance $Z_D$ and the apparent zenith distance $Z_a$ is approximated by
\begin{equation}
R = Z_D - Z_a \simeq \left(\frac{n^2-1}{2n^2}\right)\tan{Z_D},
\end{equation}
where the refractive index $n$ is a function of wavelength, pressure, temperature and humidity. Thus the angular change $R_1-R_2$, between two wavelengths is given by:
\begin{equation}
R_1 - R_2 = \left(\frac{n^2_1 - 1}{2n^2_1} - \frac{n^2_2 - 1}{2n^2_2}\right)\tan(Z_D),
\end{equation}
where $n$ is the refractive index at a given wavelength and $Z_D$ is the zenith distance. This effect is prominent at both visible wavelengths, where the refractive index varies the most strongly with wavelength, and at the smallest spatial scales. Each wavelength channel (spatial image at a given wavelength in the 3D cube) is shifted relative to an optimal wavelength, which is calculated to give equal shift on either side. The shift is made along the longest spatial dimension of the output data-cube (this assumes the user aligns the longest spatial axis of the instrument field of view along the parallactic angle). After the data-cube is rebinned to the chosen spaxel scale, the ADR is `corrected' by shifting the wavelength channels back again to emulate the correction achieved by a data reduction pipeline. This stage reduces the common field of view of the data-cube. The second order effect whereby the axis of ADR moves throughout an exposure causing the object to blur in the data-cube, is not included.

\subsection{Point spread function convolution and spatial rebinning}

Each wavelength channel is convolved with an AO spatial PSF. We have developed a novel method for parameterising AO PSFs as detailed in \citet{Zieleniewski2013} allowing us to generate PSFs at any wavelength within the HARMONI range. This is described in more detail in Section~\ref{psfs}. The pipeline generates PSFs at $1\,\rmn{mas}$ sampling and performs the PSF convolution at a scale of 1/10th of the output spatial scale in order to minimise convolution effects due to the finite sampling of the input cube and PSF. If the input data-cube is coarser than 1/10th of the output scale then it is interpolated up to 1/10th, ensuring flux conservation.

The data-cube is rebinned to the chosen output spatial scale ensuring flux conservation.

\subsection{Background and throughput}

Total background and throughput cubes are generated incorporating the effects of sky, telescope, instrument and detector. We use the ESO Skycalc sky model described by \citet{Noll2012} and \citet{Jones2013} for both emission and transmission. We model the telescope as a grey-body whereby a thermal black-body curve for the given site temperature is multiplied by a constant emissivity.

\subsection{Noise}

Poisson noise from the object, sky and telescope background, and detector dark current is added along with detector read-out noise. We use separate statistics to represent the use of CCDs for visible wavelengths and NIR arrays for longer wavelengths. The NIR detector statistics are modelled on the KMOS HAWAII-2RG detectors \citep{Finger2008} and the CCD statistics based on the MUSE calibration data (D. Ives, priv. comm.).

\subsection{Outputs}

The final outputs of {\sc hsim} are:\\
1. A mock observed cube: the main product of the simulation containing flux from the source, background and detector along with all associated noise for each $(x,y,\lambda)$ pixel of the cube;\\
2. A background cube containing all background flux;\\
3. A SNR cube giving the signal-to-noise ratio for each pixel calculated as
\begin{equation}
\label{snr_eq}
{\it SNR} = \frac{O(x,y,\lambda)T_{\rmn{exp}}\sqrt{N_{\rmn{exp}}}}{\sqrt{O(x,y,\lambda)T_{\rmn{exp}} + B(x,y,\lambda)T_{\rmn{exp}} + D\cdot T_{\rmn{exp}} + \sigma_{\rmn{R}}^2}},
\end{equation}
where $O(x,y,\lambda)$ is the object counts per second for a pixel in the 3D cube, $B(x,y,\lambda)$ is the total background counts per second, $D$ is the total dark current per second, $\sigma_{\rmn{R}}$ is the read-out noise, $T_\rmn{exp}$ is the exposure time in seconds and $N_{\rmn{exp}}$ is the total number of exposures.\\
There are also options to return a noiseless object cube containing only source flux, return a transmission cube, and perform automatic perfect sky subtraction (observed cube - background cube).

\begin{figure}
\centering
\resizebox{0.75\columnwidth}{!}{\includegraphics{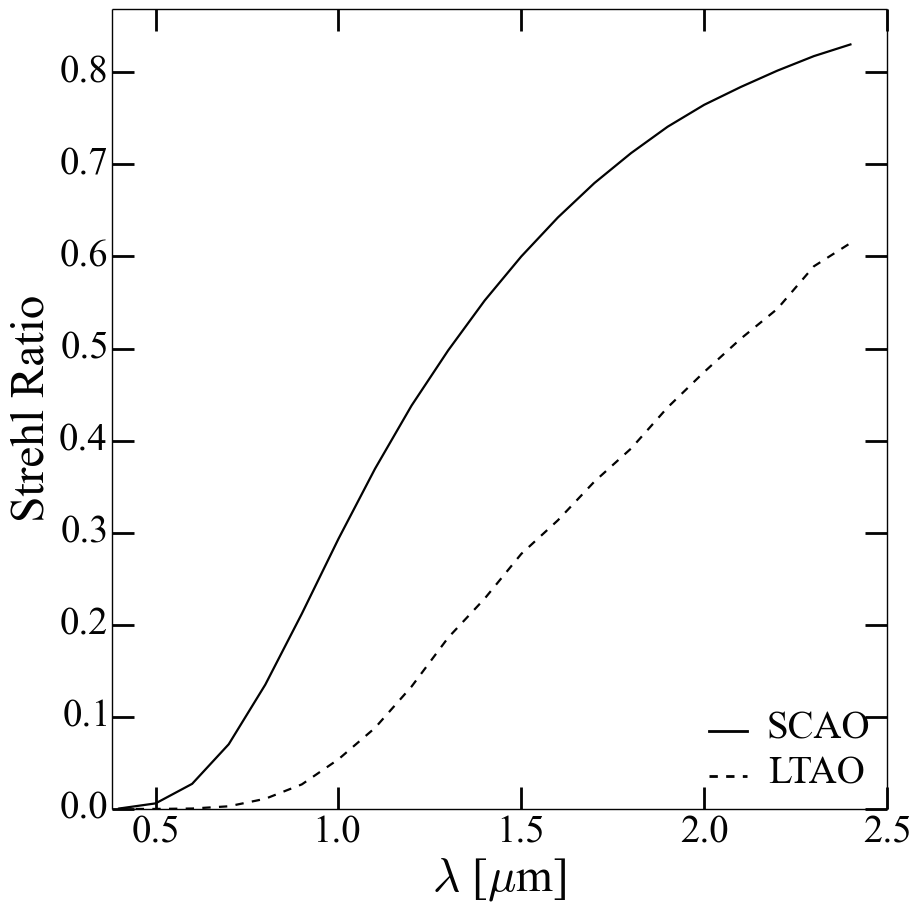}}
\caption{Strehl ratios of parameterised PSFs used in {\sc hsim} as a function of wavelength for SCAO (solid lines) and LTAO (dashed lines). PSFs are generated at a seeing of $0.67''$ and sampling of $1\,\rmn{mas}$. Both the smooth variation of AO performance and the improved performance at longer wavelengths are evident.}
\label{psf_srs}
\end{figure}

\begin{figure*}
\centering
\resizebox{15cm}{!}{\includegraphics{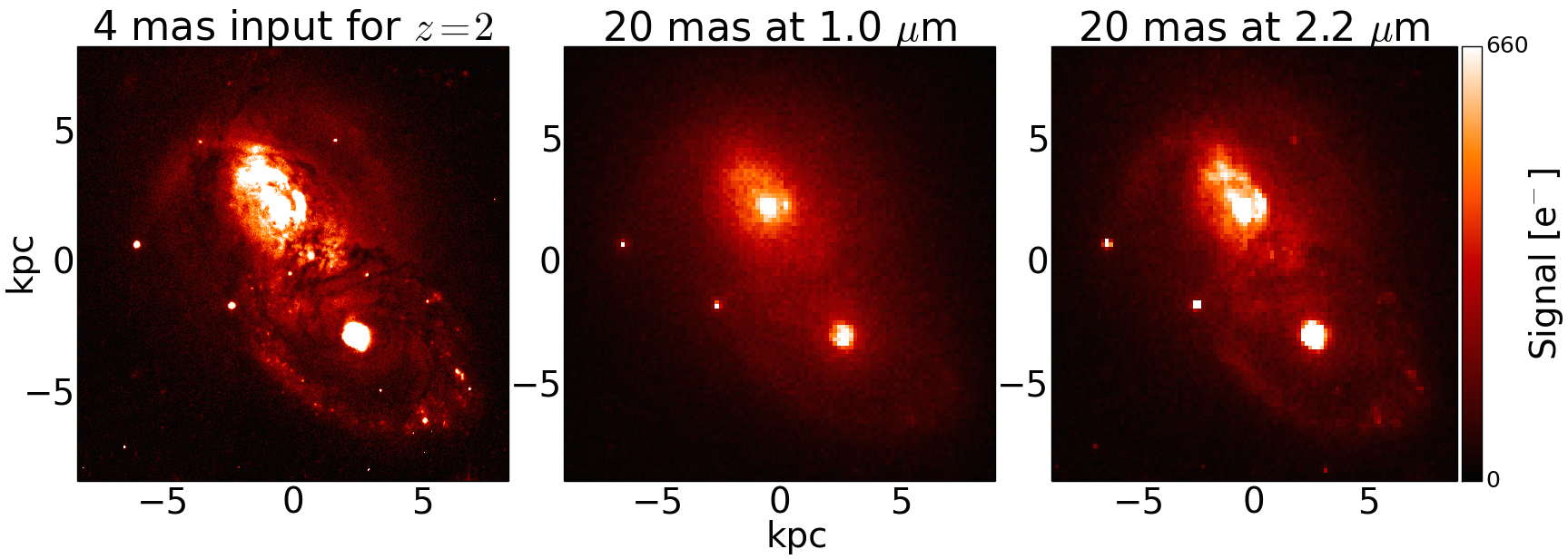}}
\caption{Integrated continuum maps from an {\sc hsim} simulation of a ULIRG observed with LTAO. Left panel shows the input map at a sampling of $4\,\rmn{mas}$ after scaling an {\it HST WCS} image of a local galaxy ($50\,\rmn{mas}$ sampling) to the correct spatial extent at $z=2$. Centre panel shows a $20\,\rmn{mas}$ continuum map at $1.0\,\rmn{\mu m}$ in the $Iz$-band. Right panel shows a continuum map at $2.2\,\rmn{\mu m}$ in the $K$-band. The observed maps are shown at ${\it SNR}=10$. We assume identical intensity and spatial distribution for both continuum maps. The improved PSF performance at longer wavelengths is clearly evident.}
\label{ulirgs}
\end{figure*}

\section[psf]{Effects of the AO Point Spread Function}
\label{psfs}

The effects of the telescope and AO system on observations are encoded in the spatial PSF. The E-ELT PSF is predicted to be a strong function of wavelength as well as other parameters including seeing, off-axis distance from reference AO star, and guide star magnitude. Its complex form means it needs to be incorporated carefully into the simulations not simply modelled as a simple Gaussian or Moffat function. We have incorporated a continuously varying AO PSF as a function of wavelength and seeing, for both LTAO and SCAO, in our simulation pipeline, using the {\it eltpsffit} program developed by J. Liske\footnote{\url{https://www.eso.org/sci/facilities/eelt/science/drm/tech_data/ao/psf_fitting/} - last accessed 01-06-15}. This allows the user to fit a 1D average radial profile of a PSF with a set of analytical functions. We used a set of simulated long-exposure LTAO and SCAO PSFs (T. Fusco \& N. Schwartz, priv. comm.), covering the HARMONI wavelength range, and fitted each of these with a combination of an obscured Airy function, Moffat function and Lorentz function \citep{Zieleniewski2013}. Interpolating the parameters of each analytical function with wavelength, we then obtain the PSF at each wavelength. We have so far only used on-axis PSFs and we assume the PSF does not vary spatially within the field of view. Fig.~\ref{psf_srs} shows the measured Strehl ratios of our parameterised PSFs for both LTAO and SCAO. The smooth variation of AO performance with wavelength in our parameterised PSFs is clear.

The importance of the wavelength variation is worth emphasising. Whilst the diffraction limit of a telescope increases from visible to NIR wavelengths, AO performance vastly improves at longer wavelengths because the atmospheric turbulence becomes relatively less disruptive. This large improvement, as seen in Fig.~\ref{psf_srs}, more than compensates for the increasing diffraction limit at longer wavelengths. We demonstrate this qualitatively in Fig. \ref{ulirgs}, which shows two continuum maps of a simulated HARMONI observation of an ultra luminous infra-red galaxy (ULIRG) at a redshift of $z=2$. The input {\it HST ACS} image at a resolution of $50\,\rmn{mas}$ \citep[IRAS06076-2139: from][]{Armus2009} has been scaled spatially to $4\,\rmn{mas}$ at $z=2$. The left map shows the object as observed in the $Iz$-band at $1.0\,\rmn{\mu m}$. The right map shows the object as observed in the $K$-band at $2.2\,\rmn{\mu m}$. For easier comparison we assume equal flux at both wavelengths and we ignored the increased thermal background in the $K$-band. It is clear from the two maps that the spatial resolution in the $K$-band is much improved over the $Iz$-band, providing exquisite detail of the underlying star-forming morphology of the ULIRG.

The AO PSFs that we used for the parameterisation also include a small amount of intrinsic jitter or blur, whicht estimates the effect of wind-shake jitter on the telescope structure. This corresponds to 2 and $3\,\rmn{mas}$ rms for the SCAO and LTAO PSFs respectively. We include the ability to add additional amounts of PSF blur to approximate reduced AO performance. We model this by convolving the AO PSF with a Gaussian of chosen rms width, so any additional blurring adds in quadrature.

\begin{figure*}
\centering
\subfigure{\includegraphics[width=6.75cm]{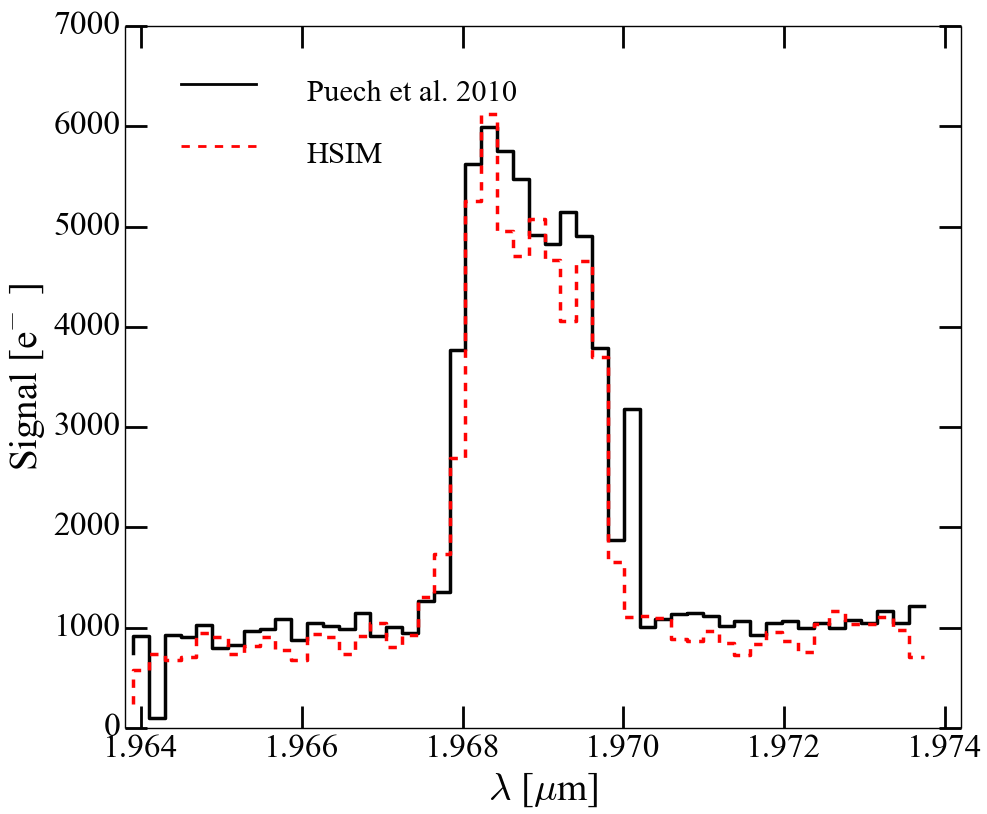}}
\subfigure{\includegraphics[width=6.5cm]{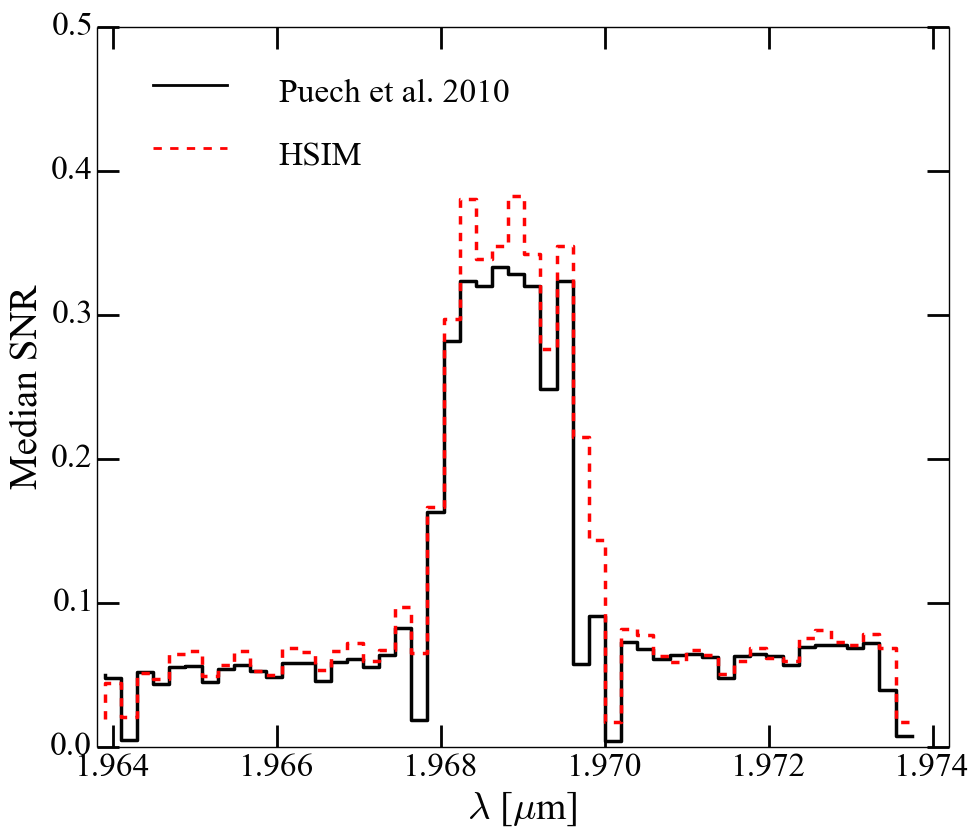}}
\caption{Left: comparison of the signal extracted from a 10 spaxel radius aperture centred on the galaxy between Puech et al. simulation and {\sc hsim}. Right: comparison of the median SNR over every spaxel in each wavelength channel. The simulations are consistent to within the noise realisations.}
\label{puechsignal}
\end{figure*}

\section{Verification Run}
\label{verification}

To test the results of {\sc hsim} we compare to two separate tools. Firstly, we compared to the pipeline of \citet{Puech2010a}, which has itself been verified using real SINFONI observations from \citet{Genzel2006}. As part of the E-ELT design reference mission (DRM) science case C10: {\it The physics and mass assembly of galaxies out to $z\sim6$}, \citeauthor{Puech2010a} undertook simulations of which some examples are publicly available\footnote{\url{https://www.eso.org/sci/facilities/eelt/science/drm/C10/} - last accessed 01-06-15}. We took the example data-cube UGC5253, a rotating disc galaxy with prominent H$\alpha$ emission.

We converted the cube for compatibility with our pipeline using the description from the DRM. We then ran simulations using identical parameters as used by \citeauthor{Puech2010a} including the identical spatial PSF. The only difference between the two runs is that Puech et al. do not specify the sky transmission spectrum they use, so we manually convolved one from the E-ELT DRM to match the spectral resolution of the data-cube.

The outputs of the comparison are shown in Fig.~\ref{puechsignal}. Our pipeline computes the total signal of $N_{\rmn{exp}}$ combined exposures, so we have divided our output cube by $N_{\rmn{exp}}$ to get a mean value for each pixel (cf. \citeauthor{Puech2010a} who determine the median value of each pixel from $N_{\rmn{exp}}$ cubes). It is clear from both the signal and SNR plots that we are computing consistent values within the uncertainty of different transmission functions, and can be confident that our pipeline is working correctly.

\begin{figure}
\centering
\resizebox{0.90\columnwidth}{!}{\includegraphics{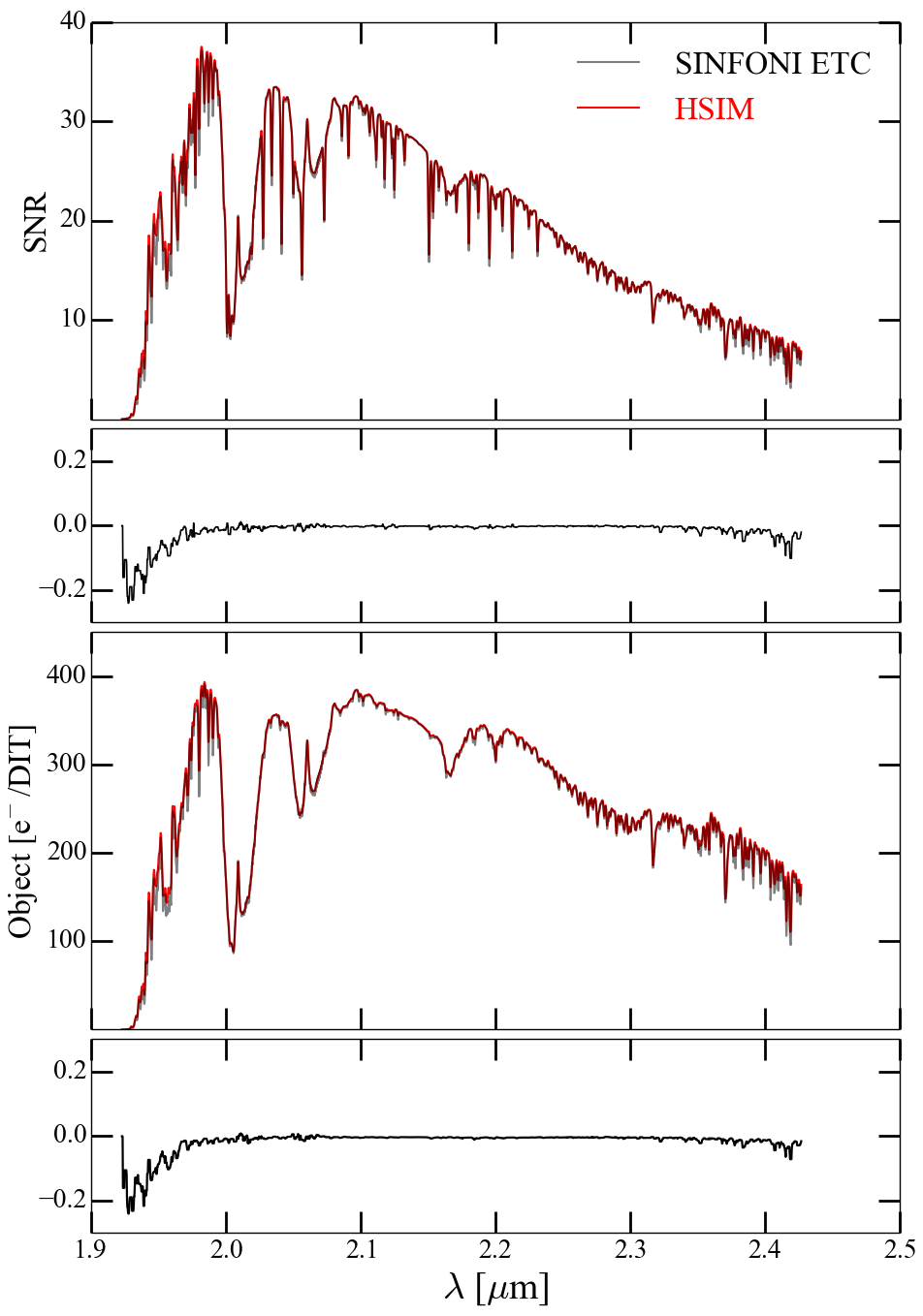}}
\caption{Comparison of the total SNR and object signal between {\sc hsim} (red) and the SINFONI ETC (grey). The two main panels show the total SNR and object signal within two spaxels. The two smaller panels show the respective residuals after median smoothing by 10 spectral pixels.}
\label{sinfonicomparison}
\end{figure}

For the second crosscheck we compared our pipeline with the ESO SINFONI exposure time calculator (ETC). We ran the ETC for an extended object with an A0V spectrum at a $K$-band Vega magnitude of 14, observed with no adaptive optics for two hours ($8\times900\,\text{s}$). The ETC returns the input spectrum which we used to create an input cube. We simulated an identical observation with {\sc hsim}, using a Gaussian PSF and the relevant VLT telescope parameters. We were careful to calculate our object flux over two spatial pixels as done by the ETC. We note that we approximated the output resolution of our spectrum using the central wavelength value of the spectrum divided by the resolving power of the SINFONI grating, as it is unclear exactly how the LSF convolution is performed in the ETC. We show the results of a comparison in Fig.~\ref{sinfonicomparison}. The two main panels show the total SNR and object signal within two spaxels. The two smaller panels show the respective residuals after median smoothing by 10 spectral pixels to account for potential differences in LSF convolution. Our results show extremely good agreement with the ETC with sub 5 per cent residuals over the majority of the spectrum (the largest residuals appearing at the blue end of the spectrum where the flux values are very small and edge effects occur). This gives us further confidence that our pipeline is working correctly and is producing realistic signal and noise values.

\section{Sensitivity Predictions}
\label{sensitivities}

In this section we present instrument point source sensitivities for several operating configurations of HARMONI as computed using {\sc hsim}. We compute the limiting AB magnitude $M_{\rmn{AB}}$ to achieve ${\it SNR}=5$ per spectral resolution element (2 pixels) after five hours on source ($T_{\rmn{exp}}=900\,\rmn{s}$). We assume sky subtraction with no penalty to on-source exposure time, using `nodding-on' or similar techniques. To do this we calculate the SNR for a given $M_{\rmn{AB}}$ after 5 hours on source, then compute the multiplicative factor $f$ required to give a $M_{\rmn{AB}}$ that yields ${\it SNR}=5$. Adding this factor (and extra factors for sky subtraction and summing over two spectral pixels) equation~\ref{snr_eq} becomes
\begin{equation}
{\it SNR} = \frac{f\cdot O\cdot T_{\rmn{exp}}\sqrt{N_\rmn{exp}}\sqrt{2}}{\sqrt{(f\cdot O + 2B + 2D)\cdot T_\rmn{exp} + 2\sigma_{\rmn{R}}^2}},
\end{equation}
and this is solved for $f$ using the standard quadratic equation where the negative solution is discarded. The SNR is calculated within a $2\times2$ spaxel extraction aperture, although for the $30\times60\,\rmn{mas}$ scale we upsize to a $120\times120\,\rmn{mas}$ box. We compare $R$, $H$ and $K$ bands. For each band we create an input cube containing a point source with a flat spectrum (constant $M_{\rmn{AB}}$). The input cubes are generated with a spatial sampling of $0.4\,\rmn{mas}$ for the $4\times4\,\rmn{mas}$ output scale and at $1\,\rmn{mas}$ for the $10\times10$, $20\times20$ and $30\times60\,\rmn{mas}$ output scales. The input cube is generated with $\sim30$ wavelength channels so as to be computationally efficient. The SNR is an average value over all the wavelength channels in the spectrum extracted from the $2\times2$ spaxel box. For each grating we choose wavelength regions minimally affected by sky lines and telluric absorption. The central wavelengths for each input cube are $0.677\,\rmn{\mu m}$ for $R$-band, $1.76\,\rmn{\mu m}$ for $H$-band, and $2.145\,\rmn{\mu m}$ for $K$-band.

These sensitivity calculations are intended to provide a general indication of the performance of the current instrument design. Whilst the instrument parameters (e.g. throughput, dark current, read-out noise) are not set at the minimum specification, we use simple $2\times2$ spaxel aperture extraction to present reasonably conservative performance estimates. Improved SNR would be obtained using the more rigorous method of optimal extraction \citep{Horne1986, Robertson1986}.

\subsection{Simulation parameters}
\label{simulation_parameters}

We set the following instrument and site parameters as to provide conservative estimates for the resulting performance.\\
{\bf Site and telescope:} We use a Paranal-like site (and telescope) temperature of $280.5\,\rmn{K}$. Seeing is set to the Cerro Armazones median value of $0.67''$ FWHM at $0.5\,\rmn{\mu m}$. We use the official $39\,\rmn{m}$ E-ELT model, which has a maximum all-glass diameter of $37\,\rmn{m}$, an obscuration ratio of 0.3 and six spider arms, giving a total collecting area of $A=932.5\,\rmn{m}^2$. We use an emissivity of 0.244 motivated by the current telescope design of seven warm reflections (5 mirror E-ELT + 2 further mirrors into HARMONI on the Nasmyth platform) of protected silver and aluminium (MgF$_\text{2}$ on Ag+Al) coated mirrors and six unmasked spider arms.\\
{\bf Instrument:} We use an instrument throughput of 35 per cent. The operating temperature of $120\,\rmn{K}$ means the instrument contributes negligible thermal background.\\
{\bf Detector model:} The NIR detector statistics are modelled on the KMOS HAWAII-2RG detectors \citep{Finger2008}. We use a dark current value of $0.0053\,\rmn{e}^-\,\rmn{s}^{-1}\,\rmn{pix}^{-1}$ and read-out noise of $2.845\,\rmn{e}^-\,\rmn{pix}^{-1}$. For the visible CCD detector statistics we adopt a dark current of $0.00042\,\rmn{e}^-\,\rmn{s}^{-1}\,\rmn{pix}^{-1}$ and read-out noise of $2.0\,\rmn{e}^-\,\rmn{pix}^{-1}$. The wavelength cut between visible and NIR detectors is set at $0.8\,\rmn{\mu m}$.

\subsection{Results}

We present the results of sensitivities for LTAO observations in tables~\ref{sensitivities:LTAO} and \ref{sensitivities:50EE}. Table~\ref{sensitivities:LTAO} gives the estimated limiting magnitudes to achieve ${\it SNR}=5$ measured from a $2\times2$ spaxel aperture ($4\times2$ for the $30\times60\,\rmn{mas}$ scale). These are separated into visible $R$-band, and NIR $H$ and $K$ bands and for each resolving power. In the $R$-band sensitivity increases with spaxel size. AO performance at visible wavelengths is poor, with Strehl ratios of $\sim1$ per cent, so larger spaxels incorporate more object flux. In the $H$-band the $20\times20\,\rmn{mas}$ scale offers slightly improved sensitivity compared to the $10\times10\,\rmn{mas}$ scale and $\sim\,1\,\rmn{mag}$ greater sensitivity compared to the $4\times4\,\rmn{mas}$ scale. However, in the $K$-band, the $10\times10\,\rmn{mas}$ is $\sim0.4\,\rmn{mag}$ more sensitive than the $20\times20\,\rmn{mas}$ scale. This is due to the increased thermal background in $K$-band, which penalises the larger spaxel scales. We also perform identical simulations for SCAO and seeing-limited observations. We find that SCAO results in $\sim0.6\,\rmn{mag}$ better sensitivity in $H$-band and $\sim0.4\,\rmn{mag}$ improvement in $K$-band. $R$-band observations with SCAO are not possible due to the visible light being used for AO correction. Seeing-limited limiting magnitudes are $\sim1.5\,\rmn{mag}$ lower than LTAO for $R$-band at the $30\times60\,\rmn{mas}$ scale and $\sim2.5\,\rmn{mag}$ lower for $H$ and $K$-bands.

\begin{figure*}
\centering
\resizebox{10cm}{!}{\includegraphics{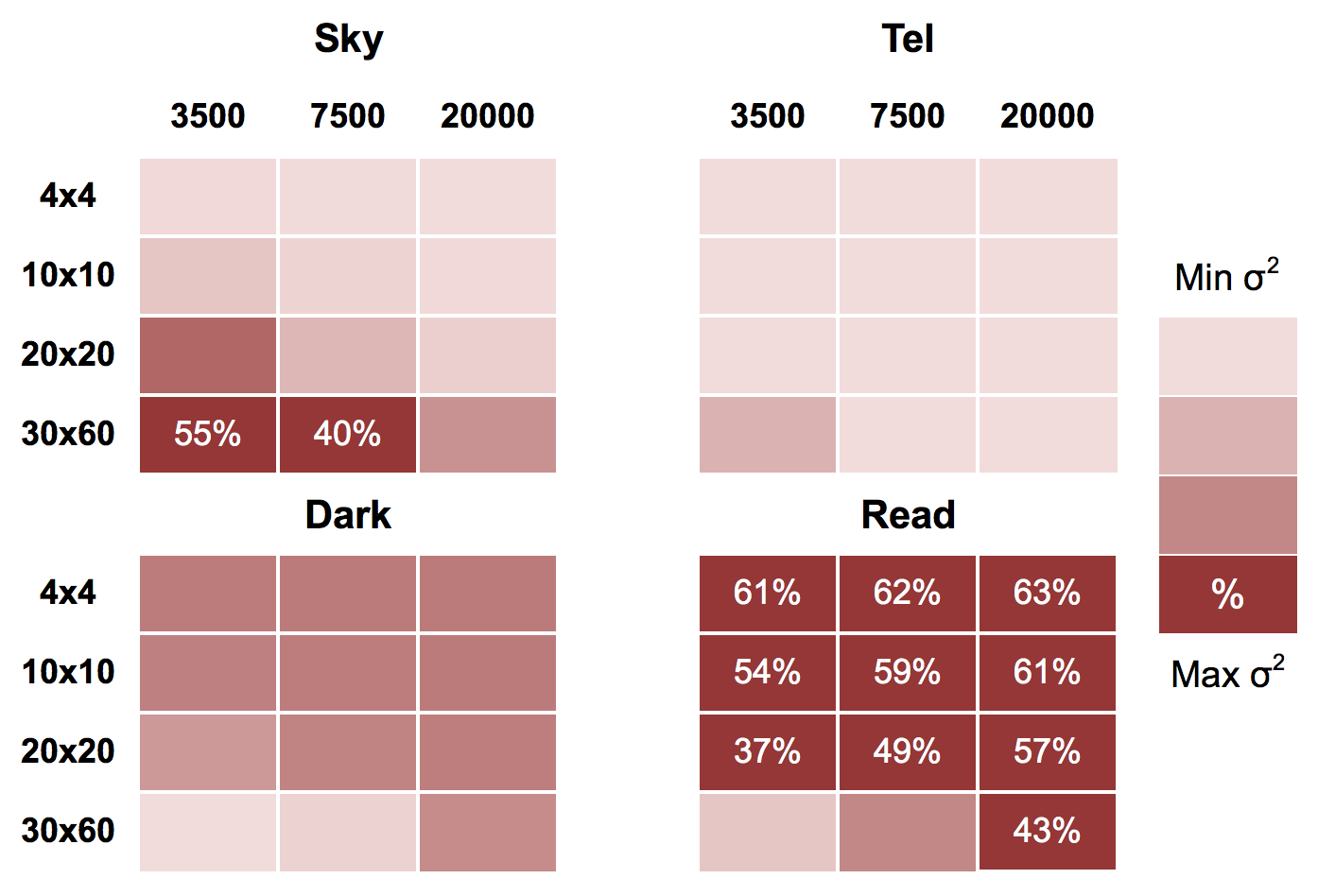}}
\caption{Heat map showing contributions to the variance for LTAO point source observations in $H$-band for five hours on source ($T_{\rmn{exp}}=900\,\rmn{s}$) after extracting in a $2\times2$ spaxel aperture centred on the object. The four quadrants show the contributions to the variance from (clockwise from top left) sky, telescope, read-out noise and dark current respectively for each observing mode of the instrument (spaxel scale and resolving power). The instrument thermal background is assumed to be negligible. Dark colours show a greater contribution and light colours show a smaller contribution. For each instrument configuration, the box with the greatest contribution contains its percentage contribution to the total variance. Observations in $H$-band are predominately read-out noise limited except at the coarsest spaxel scale and lowest resolving powers which are sky limited.}
\label{LTAO_H-band}
\end{figure*}

\begin{figure*}
\centering
\resizebox{10cm}{!}{\includegraphics{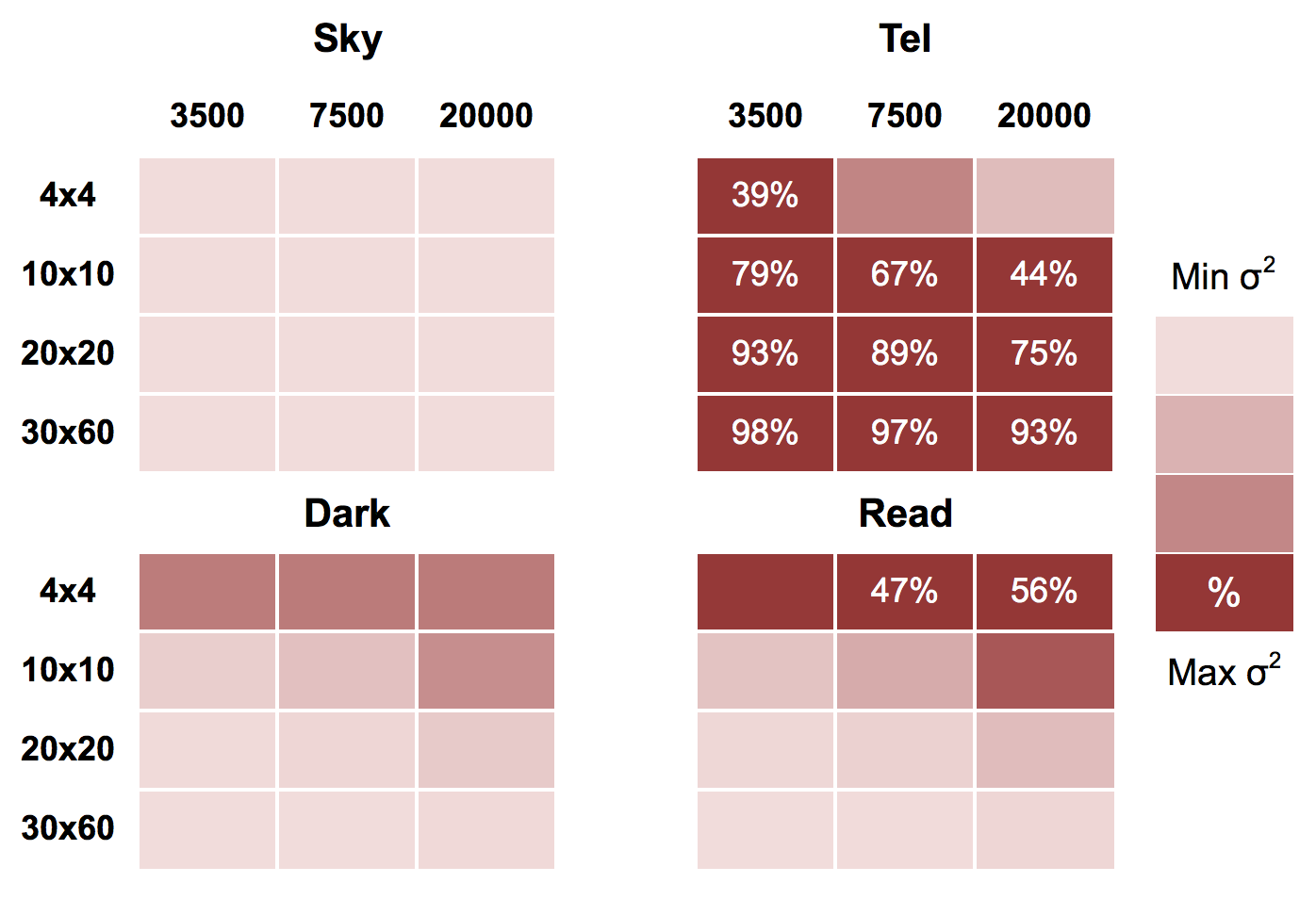}}
\caption{Same as Fig.~\ref{LTAO_H-band} but for $K$-band observations: heat map showing contributions to the variance for LTAO point source observations in $K$-band for five hours on source ($T_{\rmn{exp}}=900\,\rmn{s}$) after extracting in a $2\times2$ spaxel aperture centred on the object. Comparing to Fig.~\ref{LTAO_H-band} the $K$-band observations are dominated by the telescope thermal background, except at finest spaxel scale and highest resolving powers which are read-out noise limited.}
\label{LTAO_K-band}
\end{figure*}

Table~\ref{sensitivities:50EE} shows magnitudes for identical observations as table~\ref{sensitivities:LTAO} but after extracting the spectrum from an aperture containing 50 per cent of the ensquared energy of the PSF. For the $R$, $H$ and $K$ bands these correspond to $400\times400$, $140\times140$ and $80\times80\,\rmn{mas}$ respectively. We omit values for the $30\times60\,\rmn{mas}$ spaxel scale in $K$-band because the $4\times2$ spaxel aperture already contains greater than 50 per cent of the ensquared energy. This table shows that a greater SNR can be achieved in the $R$-band by increasing the aperture size. For the $4\times4$ and $10\times10\,\rmn{mas}$ scales, a factor of $\sim1\,\rmn{mag}$ improvement is gained. However, for both $H$ and $K$ bands, the limiting magnitudes are lower because the extra spaxels from the larger extraction aperture contribute more detector noise, as can be seen from Figs~\ref{LTAO_H-band} and \ref{LTAO_K-band}.

\begin{table}
\caption{Point source sensitivity predictions for HARMONI with LTAO calculated from a $2\times2$ spaxel aperture centred on the object.}
\label{sensitivities:LTAO}
  \begin{tabular}{p{1.3cm} p{1.3cm} r r r r r r}
  \multicolumn{6}{c}{{\bf ${\bf 2\times2}$ spaxel aperture}}\\
  \hline
 Band & R  & $4\times4$ & $10\times10$ & $20\times20$ & $30\times60$ \\
  \hline
 $R$ & 3500 & 22.7 & 23.6 & 24.3 & 25.1\vspace{5pt}\\
 $H$ & 3500 & 26.2 & 27.0 & 27.0 & 26.4 \\
 & 7500 & 25.2 & 26.0 & 26.1 & 25.7 \\
   \vspace{5pt}
 & 20000 & 24.1 & 25.0 & 25.1 & 24.9 \\
 $K$ & 3500 & 25.7 & 26.3 & 25.8 & 24.8 \\
 & 7500 & 25.2 & 25.8 & 25.4 & 24.5 \\
 & 20000 & 24.2 & 25.1 & 24.8 & 23.9 \\
\hline
\multicolumn{6}{l}{\begin{minipage}{\columnwidth}Notes: Limiting AB magnitudes to achieve ${\it SNR}=5$ per spectral resolution element for five hours on-source ($T_{\rmn{exp}}=900\,\rmn{s}$). We assume sky subtraction with no penalty to on-source exposure time. Band column gives the bandpass region, R is the resolving power and the remaining columns give the limiting magnitude for each HARMONI spaxel scale. $4\times2$ spaxel aperture for $30\times60\,\rmn{mas}$ scale.\end{minipage}}\\
\end{tabular}
\end{table}

\begin{table}
  \caption{Point source sensitivity predictions for HARMONI with LTAO calculated from a square aperture that encloses 50 per cent of the ensquared energy (EE) for each band.}
  \label{sensitivities:50EE}
\begin{tabular}{p{1.3cm} p{1.3cm} r r r r r r}
  \multicolumn{6}{c}{{\bf 50 per cent EE square aperture}}\\
  \hline
 Band & R  & $4\times4$ & $10\times10$ & $20\times20$ & $30\times60$ \\
  \hline
 $R$ & 3500 & 23.6 & 24.6 & 25.1 & 25.4\vspace{5pt}\\
 $H$ & 3500 & 24.5 & 25.4 & 26.0 & 26.2 \\
 & 7500 & 23.5 & 24.5 & 25.1 & 25.6 \\
 \vspace{5pt}
 & 20000 & 22.5 & 23.4 & 24.1 & 24.7 \\
 $K$ & 3500 & 24.7 & 25.1 & 25.1 & - \\
 & 7500 & 24.1 & 24.6 & 24.8 & - \\
 & 20000 & 23.1 & 23.9 & 24.2 & - \\
\hline
\multicolumn{6}{c}{\begin{minipage}{\columnwidth}Notes: Limiting AB magnitudes to achieve ${\it SNR}=5$ per spectral resolution element for five hours on-source ($T_{\rmn{exp}}=900\,\rmn{s}$). For the $R$, $H$ and $K$ bands the 50 per cent EE apertures correspond to $400\times400$, $140\times140$ and $80\times80\,\rmn{mas}$ respectively. We assume sky subtraction with no penalty to on-source exposure time. Band column gives the bandpass region, R is the resolving power and the remaining columns give the limiting magnitude for each HARMONI spaxel scale.\end{minipage}}\\
\end{tabular}
\end{table}

To illustrate the different noise regimes,  Figs~\ref{LTAO_H-band} and \ref{LTAO_K-band} show heat maps representing the contributions to the total variance from each source: sky, telescope, dark current, and read-out noise, when observing with LTAO in the $H$-band and $K$-band respectively. Each observing configuration is represented by a single box. Light colours show small contribution and dark colours show larger contribution. For each observing configuration, the box with the largest contributor contains the percentage of the total variance from that source. It is clear from these figures that in $H$-band HARMONI is generally detector read-out noise dominated, except for the configuration of $30\times60\,\rmn{mas}$ scale combined with the two lowest resolving powers, which are sky dominated. However, in the $K$-band observations are telescope (thermal) background dominated, except for the finest spaxel scale and highest resolving powers.

\section{Case study: simulations of Emission-line Galaxies at $z\sim2-3$}
\label{swinbank}

One of the major science goals of HARMONI is to spatially resolve the
interstellar medium (ISM) of high-redshift ($z\sim$\,2--5) galaxies and measure the physical
processes occurring on scales of individual H{\sc ii} regions. At
these redshifts, the comoving star-formation density was substantially
higher than at $z\sim$\,0, and so this era has been heralded as the
peak epoch of galaxy formation when most of todays massive galaxies
formed the bulk of their stellar mass \citep{Madau2014}. At these early times, the Hubble
sequence was not in place, and galaxies appear to be undergoing
significant changes to their morphologies and stellar populations.
Over the past decade major advances have been made in measuring the
dynamics of galaxies at this epoch using integral field spectroscopy
(e.g. SINS: \citealt{Genzel2006}, \citealt{ForsterSchreiber2006, ForsterSchreiber2009}; OSIRIS: \citealt{Law2007, Law2009}; KMOS: \citealt{Wisnioski2015}; see also \citealt{Glazebrook2013} review).  However, due to the photon-starved nature of the
observations, deriving the dynamics of high redshift galaxies requires
long integration times, and only offers limited spatial resolution.
Moreover, even with natural- or laser-guide star AO
assisted observations, spatially resolved studies are limited to
$\gtrsim1\,\rmn{kpc}$ resolution \citep[e.g.][]{ForsterSchreiber2011a,
  ForsterSchreiber2011b, Swinbank2012a, Swinbank2012b}. Gravitationally lensed star-forming galaxies do provide the unique ability to resolve regions at scales of $\sim$60--200 pc \citep{Jones2010, Livermore2012, Livermore2015}, however these are notable rare exceptions limited to a handful of objects.

The increased light grasp and spatial resolution of the E-ELT should
allow the study of high-redshift galaxies on the scales of individual
star forming H{\sc ii} regions (down to 100 pc scales).  The
properties (e.g.\ sizes, luminosities, velocity dispersions, chemical
make-up, and spatial distribution) of H{\sc ii} regions reflect the
underlying ISM (such as gas density and pressure), which in turn
reflect the dominant route by which galaxies accrete the bulk of their
gas.

The goals of the simulations presented in this section are to
investigate the ability to detect and measure kinematics of
emission-line galaxies at redshifts of $z\sim$\,2--3, using HARMONI on
the E-ELT. Our goals are to:\\
{\it (i)} determine how well the global kinematics (e.g. rotation
curves) can be derived as a function of star-formation rate, size and
morphology for galaxies at $z\sim$\,2--3;\\
{\it (ii)} determine the smallest physical scales for which physical
properties can be derived, including identifying\,/\,measuring the
properties of individual star forming regions.

To demonstrate the use of our simulation pipeline in this area, we
present results from a set of simulated observations of a sample
of emission-line galaxies at $z\sim$\,2--3. We focus on galaxies with
prominent H$\alpha$ emission, which falls into the $K$-band at these
redshifts. We highlight that our simulations are not designed to test galaxy formation models. Rather they are designed to test how well measured properties (e.g. rotation curves, clump properties) can be derived for a given exposure time (or at a given spatial resolution) using a reasonable input galaxy image/spectrum with a set of disc and clump scaling relations.

In the rest of this section, we discuss how we construct mock
data-cubes, our simulation runs, analysis methods and derived
conclusions for this science case.

\begin{figure*}
\centering
\resizebox{17cm}{!}{\includegraphics{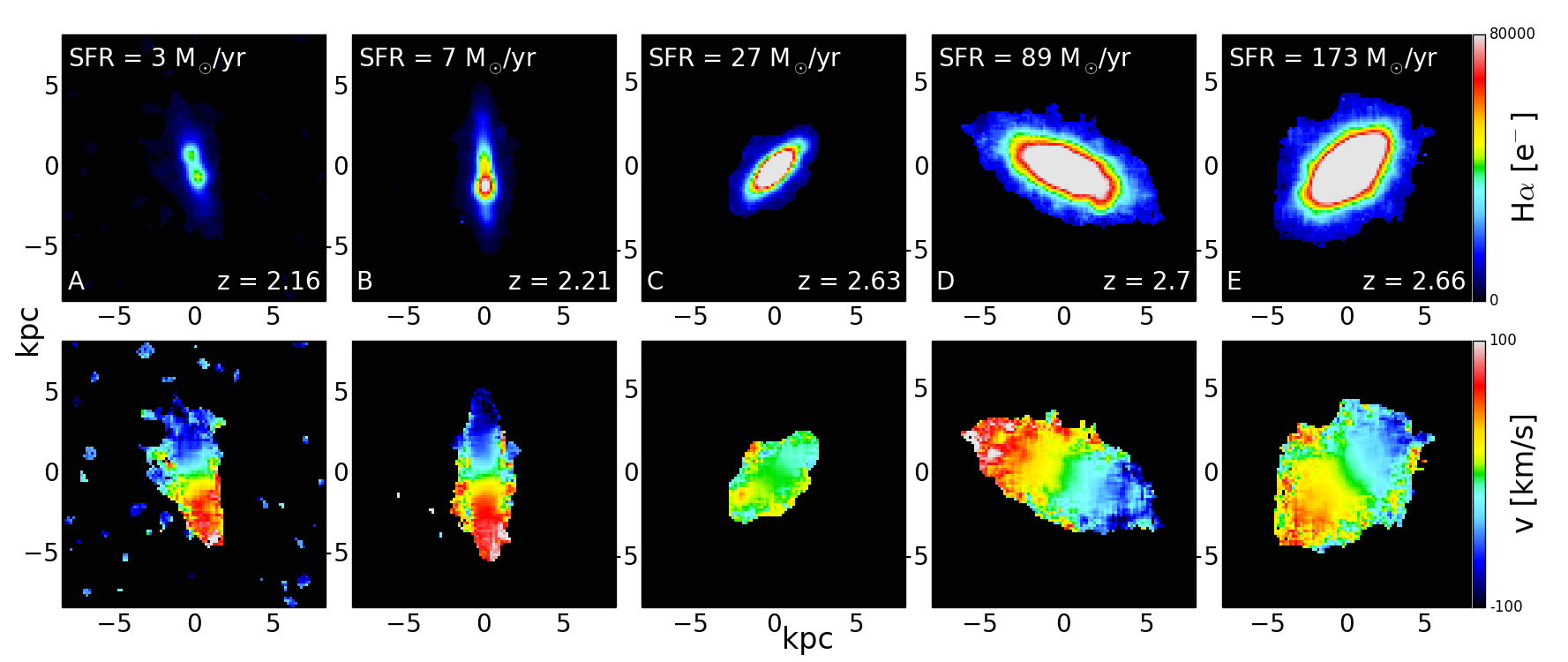}}
\caption{Maps of H$\alpha$ flux (top row) and line-of-sight velocity (bottom row) for smooth disc
  galaxies. Galaxies are ordered in increasing SFR from left to
  right. Also shown is the redshift of each galaxy. All galaxies are
  observed for 10 hours at the $20\times20\,\rmn{mas}$ ($\sim200\,\rmn{pc}$) scale. Velocity
  gradients are easily measured even for the lowest SFRs with a factor
  of $\sim5$ improvement over existing instruments in the highest SNR regions.}
\label{swinbank_galaxy_outputdiscs}
\end{figure*}

\begin{figure*}
\centering
\resizebox{16cm}{!}{\includegraphics{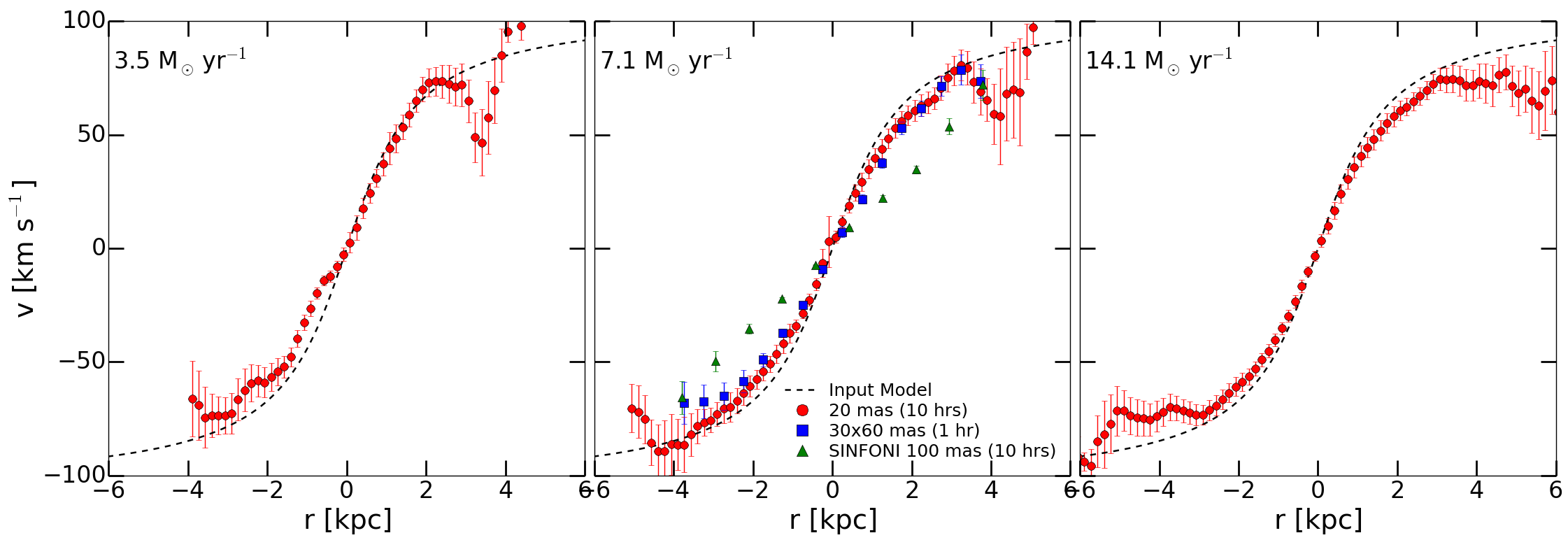}}
\caption{Rotation curves for smooth disc galaxy B with varying SFR (increasing left to right) at $20\,\rmn{mas}$ scale (red circles) as observed by the E-ELT with HARMONI. The input model curve is shown as the dashed black line. Also plotted in the centre subplot is the rotation curve for a 1 hour observation at the $30\times60\,\rmn{mas}$ scale (blue squares), which has been binned up to 60 mas spaxels. We also show a simulated 10 hour VLT (SINFONI) observation at $100\,\rmn{mas}$ (green triangles). All curves are extracted along a $\sim1\,\rmn{kpc}$ slit aligned along the semi-major axis. From these simulations the E-ELT with HARMONI offers a 10 fold improvement in observing efficiency compared to current telescopes, i.e. the E-ELT could observe 10 times as many objects in equal observing times. It also offers finer resolution, with 5$\times$ more independent data points at the 20 mas scale, and improved sensitivity for equal observing times on the same object}
\label{swinbank_galaxy_rotcurves}
\end{figure*}

\subsection{Input data-cubes}

We generate a sample of input galaxy data-cubes covering a range of
star formation rates, redshifts and morphologies ranging from smooth
exponential discs of gas and stars to `clumpier' galaxies (gas follows stars in each case).  Each galaxy
datacube is generated with values randomly picked from a range of uniformly distributed physical parameters: redshift (2.05--2.8), total disc star formation rate SFR (1--200$\,\rmn{M}_{\odot}\,\rmn{yr}^{-1}$), gas
fraction (0.1--0.9), inclination (20--70\,$\rmn{deg}$), position angle
(0--360$\,\rmn{deg}$), reddening $A_v$ (0--1.5\,$\rmn{mag}$),
half-light radius $r_{\rmn{hl}}$ (0.5--2.5\,$\rmn{kpc}$), disc
intrinsic velocity dispersion (15--40\,$\rmn{km}\,\rmn{s}^{-1}$) and metallicity
($Z$\,=\,0.05--1\,$Z_{\odot}$).  The underlying light profile of the
galaxy disc follows an exponential profile, and the velocity field
follows a simple {\it arctan} model \citep{Courteau1997}. We use the star formation law of \citet{Kennicutt1998b}. We also add a number of star-forming
regions using scaling relations inferred from
observations of lensed star-forming galaxies at $z\sim$\,1--3 \citep[e.g.\ ][]{Jones2010, Livermore2012, Livermore2015}. The number of star-forming clumps is set using the redshift dependent clump luminosity function from \citet{Livermore2012, Livermore2015}, where the normalisation is a function of the disc gas fraction and ranges from 0.01--2. The clump velocity dispersion and sizes use the scaling relations from \citet{Livermore2015}.

Finally, we include stellar continuum assuming either a constant or
exponential star formation history (with an integral that matches the
dynamical mass after accounting for gas fraction). We use a solar
metallicity simple stellar population model with a \citet{Chabrier2003} initial
mass function to derive the stellar continuum and assign this to the
disc according to its luminosity profile.

For the purposes of this analysis, we generate input cubes with a spatial
sampling of $10\times10\,\rmn{mas}$ and a resolving power of
$R$\,=\,10,000.

\subsection{Simulation runs}

We simulate a series of mock HARMONI observations of these galaxies
using LTAO.  We adopt $T_{\rmn{exp}}=900\,\rmn{s}$ and use
the $R\sim$\,3500 $H+K$ grating. We focus on simulations at
$20\times20\,\rmn{mas}$ ($\sim200\,\rmn{pc}$ at $z=$\,2--3) scales unless otherwise stated. The pipeline
parameters (site, telescope, instrument and detector) are all set
identically to those used for the sensitivity calculations (see
section~\ref{simulation_parameters}).

\subsection{Global kinematic measurements}

In Fig.~\ref{swinbank_galaxy_outputdiscs} we show the recovered
H$\alpha$ flux distribution and gas kinematics for a sample of five smooth-disc galaxies assuming a 10 hour integration. We use a Gaussian fitting routine to fit the H$\alpha$ (and
N{\sc ii}) emission lines spaxel-by-spaxel. This routine iterates over
each spaxel and fits both the continuum and Gaussian profiles to the
spectrum.  In cases where no fit is made we average over the
surrounding spaxels to increase signal at the expense of spatial
resolution. For the velocity maps, which show the global gas kinematics, we bin over a $3\times3$ spaxel box, giving a varying resolution of $20\,\rmn{mas}$ in the bright regions to $60\,\rmn{mas}$ in fainter regions. We use a SNR threshold of 7 for detection of an emission line.

From Fig.~\ref{swinbank_galaxy_outputdiscs} we see that HARMONI
is capable of measuring velocity profiles in galaxies down to Milky
Way-like star-formation rates on scales of at least $\sim200\,\rmn{pc}$ (in the brightest/highest SNR regions) in this integration time. In Fig.~\ref{swinbank_galaxy_rotcurves} we show the rotation curves for
galaxy B (extracted along a $\sim1\,\rmn{kpc}$ wide slit
aligned along the semi-major axis) for the same galaxy
properties, but with a star-formation rate of 3.5, 7 and
$14\,\rmn{M}_{\odot}$\,\,yr$^{-1}$ (increasing left to right). This galaxy has the steepest
rotation velocity from our sample so gives the best indication of
which regions can be recovered for a small range of quiescent star-formation rates. It also shows the effect limited spatial resolution can have on tracing the inner part of the rotation curve (`beam smearing').
The velocity profile is traced very closely by HARMONI at 20 mas sampling through the central part
of the galaxy for all SFRs. The half-light radius of this galaxy is $r_{\rmn{hl}}=1.3\,\rmn{kpc}$ so the curve is recovered out to $3r_{\rmn{hl}}$ for an SFR of $3.5\,\rmn{M}_{\odot}\,\rmn{yr}^{-1}$, which increases to $4.6r_{\rmn{hl}}$ for a four-fold increase in SFR.

In Fig.~\ref{swinbank_galaxy_rotcurves} we also show the rotation curve for a 1 hour E-ELT (HARMONI) observation at the coarser $30\times60\,\rmn{mas}$ scale (blue squares), which has been binned up to 60 mas spaxels. Comparing this to the rotation curve derived from a simulated 10 hour VLT (SINFONI) observation at $100\,\rmn{mas}$ we see that the E-ELT offers higher resolution data at better sampling with 10$\times$ greater efficiency in observing time. For equal observing time the VLT curve underestimates the true curve at all radii and also only extends to $3r_{\rmn{hl}}$, compared with $4.2r_{\rmn{hl}}$ from HARMONI at 20 mas sampling. Deriving a simple dynamical mass estimate from each observation gives $\sim9\times10^9\,\rmn{M}_{\odot}$ from E-ELT (HARMONI) and $\sim4\times10^9\,\rmn{M}_{\odot}$ from the VLT (SINFONI) simulation. Comparing to the input value of $8.6\times10^9\,\rmn{M}_{\odot}$ this represents both an accurate estimate and a factor of 2 improvement by HARMONI after 10 hours observing. From these simulations we find that the E-ELT with HARMONI offers a 10 fold improvement in observing efficiency compared to current telescopes, i.e. the E-ELT could observe 10 times as many objects in equal observing times. It also offers improved sensitivity and finer resolution, with 5$\times$ more independent data points at the 20 mas scale, for equal observing times on the same object.

\subsection{Detailed kinematics}

\begin{figure*}
\centering
\resizebox{17cm}{!}{\includegraphics{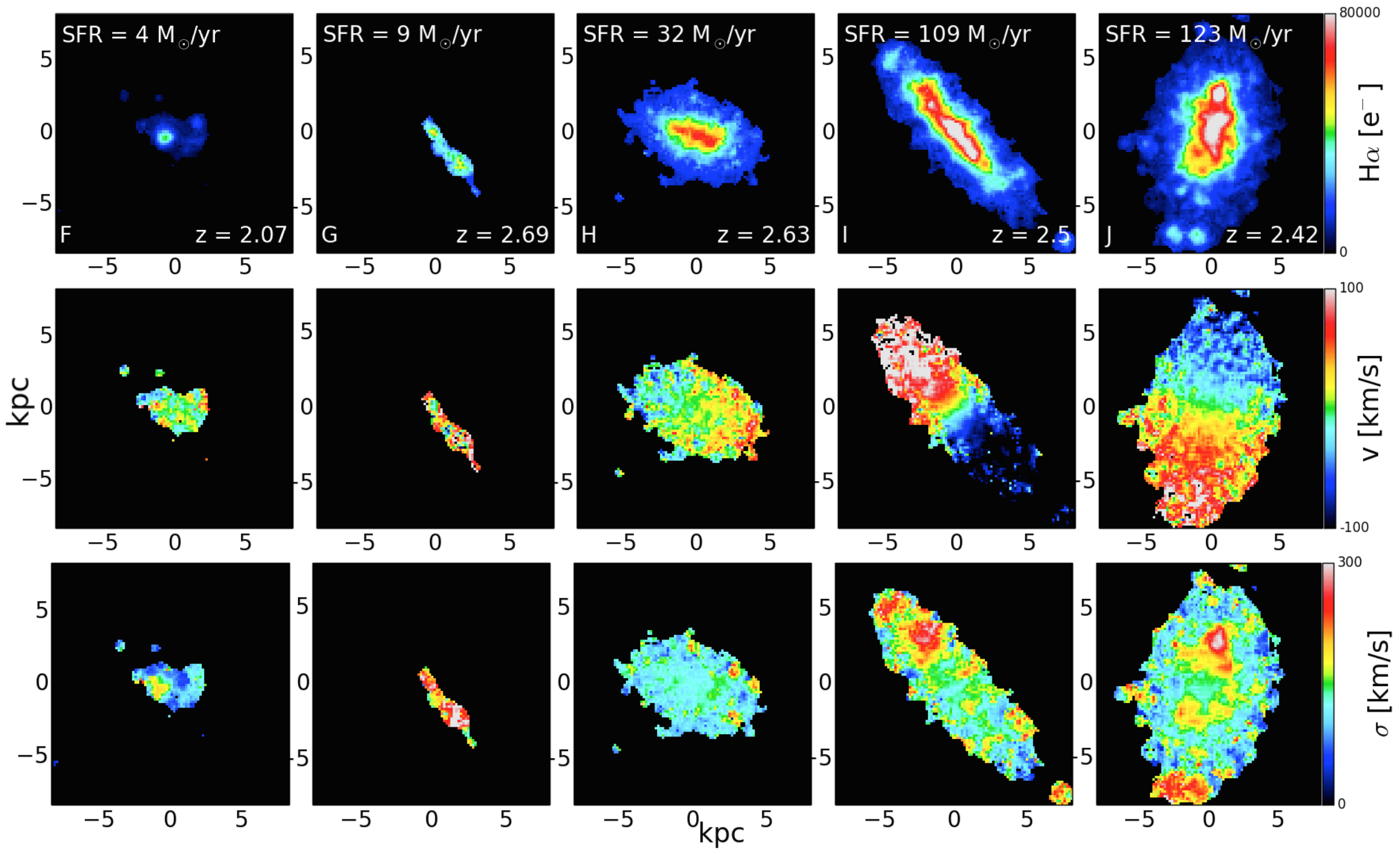}}
\caption{Maps of H$\alpha$ flux, line-of-sight velocity, and velocity dispersion for clumpy galaxies. Galaxies are ordered in increasing SFR from left to right. Also shown is the redshift of each galaxy. All galaxies are observed for 10 hours at the $20\times20\,\rmn{mas}$ scale. Velocity structure is evident even at the lowest SFRs and the velocity dispersion maps show internal structure of $\sim5$-$6$ spaxels diameter ($\sim500\,\rmn{pc}$ radius) for the faintest galaxy down to $\sim3$ spaxels ($\sim250\,\rmn{pc}$ radius) for the brightest galaxy.}
\label{swinbank_galaxy_outputclumps}
\end{figure*}

The fine spatial sampling of HARMONI coupled with LTAO will allow for
very detailed observations of $z\sim2$ galaxies. In
Fig.~\ref{swinbank_galaxy_outputclumps} we show the observed H$\alpha$
intensity and velocity dispersion maps for the clumpy galaxies in our
simulations. To maintain the high spatial resolution required for detecting individual star forming regions we bin over a $2\times2$ spaxel box where no fit is made to the emission lines, giving an effective resolution of $40\,\rmn{mas}$ ($\sim300\,\rmn{pc}$) in fainter regions, and again use a SNR
threshold of 7 for detection of an emission line. As
Fig.~\ref{swinbank_galaxy_outputclumps} shows, detailed structure is seen in both the H$\alpha$ emission maps and $\sigma$ maps. The maps of galaxy J in Fig.~\ref{swinbank_galaxy_outputclumps} (far right) show structure of $\sim3$ spaxels in diameter. Thus HARMONI will be able to make very detailed
measurements of galaxy substructure, even for galaxies with Milky Way
star-formation rates.

\begin{figure*}
\centering
\resizebox{15cm}{!}{\includegraphics{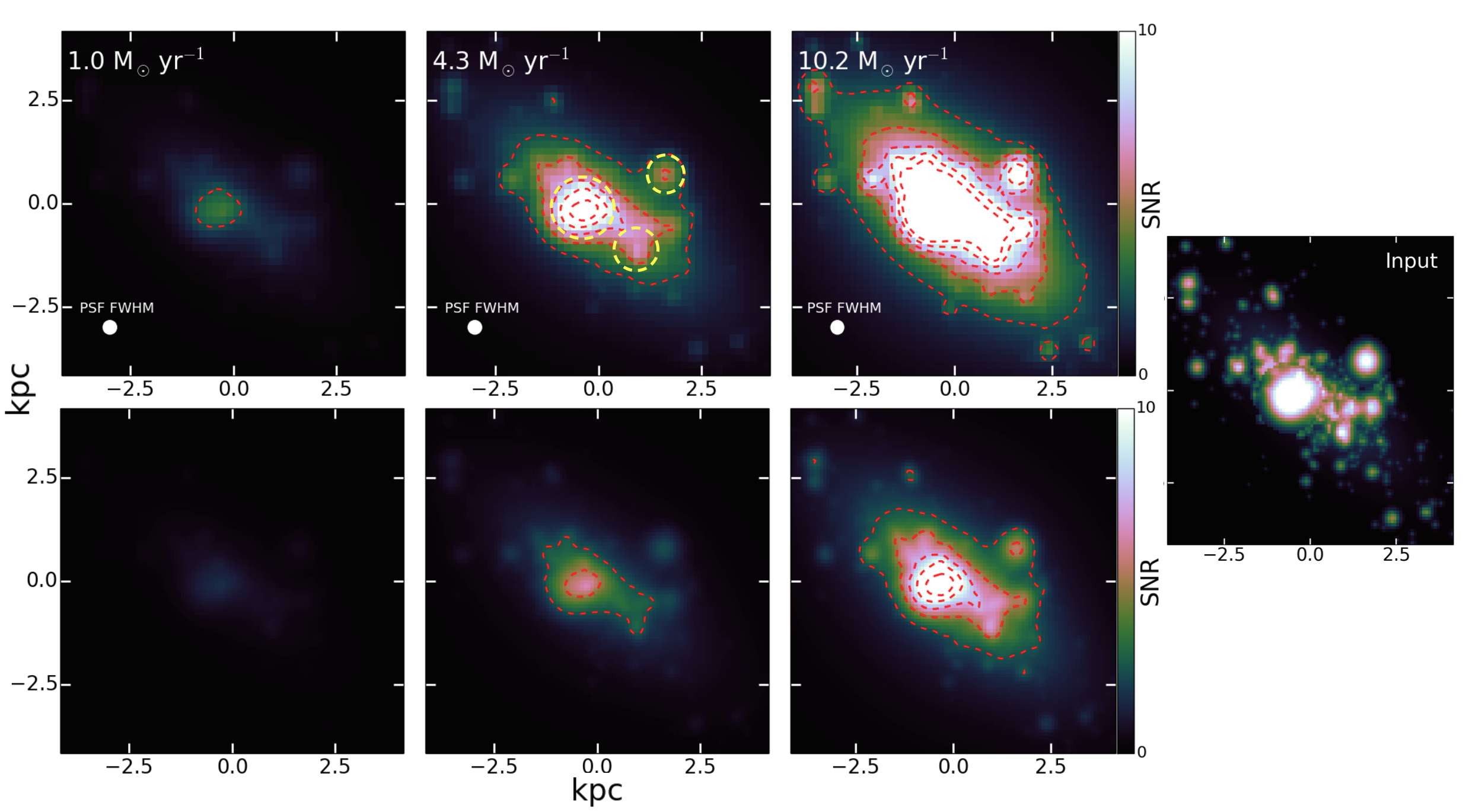}}
\caption{SNR maps of a model clumpy galaxy for varying SFRs observed
  at $20\times20\,\rmn{mas}$ (top row) and $10\times10\,\rmn{mas}$
  (bottom row). SFR increases from left to right and is denoted in the
  top left of each panel. The contours show constant SNR increasing
  from 2.5 to 12.5 in steps of 2.5. The size of the LTAO PSF is shown
  in the bottom left of the each map in the top row. The total
  exposure time is 10 hours. The three star-forming clumps detected by {\sc clumpfind} are highlighted with yellow dashed circles in the central panel of the top row. For comparison we also show the input morphology (10 mas sampling) in the right hand subplot. Sub-kpc sized star forming regions are detected with $SNR>5$ for $SFR\sim4\,\rmn{M}_{\odot}$ at $20\rmn{mas}$. Similar regions are detected for $SFR\sim10\,\rmn{M}_{\odot}$ at $10\rmn{mas}$.}
\label{clumpy_snrs}
\end{figure*}

To demonstrate HARMONI's ability to discern properties of individual
star forming regions, we focus on the galaxy with the lowest
star-formation rates from the `clumpy' galaxy simulations
(galaxy F in Fig.~\ref{swinbank_galaxy_outputclumps}). Fig.~\ref{clumpy_snrs}
shows the recovered H$\alpha$ signal-to-noise maps for a 10 hour
observation at the $20\times20\,\rmn{mas}$ (top row) and
$10\times10\,\rmn{mas}$ (bottom row) scales. For star-formation rates
below $10\,\rmn{M}_{\odot}\,\rmn{yr}^{-1}$ the $20\,\rmn{mas}$ scale
offers superior signal-to-noise, while still resolving individual
clumps.  However, at a star-formation rate of
$1\,\rmn{M}_{\odot}\,\rmn{yr}^{-1}$ no clumps are detected with a
${\it SNR} > 5$ at the $20\,\rmn{mas}$ scale in 10 hours.

To quantify the likely detection and properties of the clumps in a
HARMONI observation we use the 2D {\sc clumpfind} routine by
\citet{Williams1994} to determine positions and sizes of individual clumps. This
method has been used previously by \citet{Livermore2012,
  Livermore2015} on observations of lensed galaxies. The routine uses
isophotes to define clumps starting in the brightest regions and then
moving down through the isophote levels.  Any isolated contours are
defined as new clumps, and any which enclose an existing peak are
allocated to that clump.  A contour which encloses two or more existing
peaks has its pixels divided between them using a `friends-of-friends'
algorithm.  We follow a similar procedure as used in
\citet{Livermore2015} and set the minimum threshold at $3\sigma$ and
move up in $1\sigma$ steps.

In galaxy F we detect three clumps (indicated by yellow circles on Fig.~\ref{clumpy_snrs}), and proceed to calculate their luminosities and then star-formation rates by summing the pixels
within the clump from the H$\alpha$ map (corrected for local
background from the underlying disc). We measure star-formation rates in these three clumps to be
$1.46\pm0.02$, $0.26\pm0.01$ and
$0.23\pm0.01\,\rmn{M}_{\odot}\,\rmn{yr}^{-1}$, where the uncertainties
are derived from the H$\alpha$ variance map.  As a crosscheck of our method we integrate the complete observed H$\alpha$ map and
calculate the total star-formation rate of the galaxy to be
$4.12\pm0.03\,\rmn{M}_{\odot}\,\rmn{yr}^{-1}$ (which is similar
to the input value of $4.35\,\rmn{M}_{\odot}\,\rmn{yr}^{-1}$).

Finally, assuming the clumps have circular symmetry, we infer a radius
of each clump.  We measure the radii as $980\pm150$, $500\pm200$ and
$520\pm250\,\rmn{pc}$ respectively. These match closely to the input clump sizes of 1000, 630 and 500$\,\rmn{pc}$ respectively. Comparing these measurements to observations of lensed
galaxies (Fig.~9 of \citealt{Livermore2015}) we see that HARMONI will
be capable of detecting and measuring properties of clumps at least a factor 2 smaller than currently possible for normal (unlensed) galaxies at $z\sim2$
($\sim500\,\rmn{pc}$ compared to $\sim1\,\rmn{kpc}$). Thus, it should
be possible to observe the same galaxy at the $10\,\rmn{mas}$ scale
for a greater number of hours and measure properties of even smaller
clumps. In fact, from the signal-to-noise map of galaxy F with $SFR = 10.2\,\rmn{M}_{\odot}\,\rmn{yr}^{-1}$ in
Fig.~\ref{clumpy_snrs}, we see there are clumps of three pixels
diameter in the $20\,\rmn{mas}$ scale with ${\it SNR}>5$. This
corresponds to $\sim 250\,\rmn{pc}$. Our current analysis is a first
step to showcase the capabilities of the simulation pipeline, and we
will undertake a more thorough analysis of HARMONI's ability to detect
and measure star forming clumps in a follow-up paper.

\section{Conclusions}
\label{conclusion}

We present {\sc hsim}: a new simulation pipeline for the HARMONI integral field spectrograph on the E-ELT. The pipeline takes input data-cubes and simulates observations, folding in sky, telescope, instrument and detector parameters to create output mock data. {\sc hsim} is able to provide quantitive measures of the precision with which we can derive a number of key physical parameters for particular science cases. It allows the user to gain an understanding of the uncertainties associated with making a specific astrophysical measurement. We have described {\sc hsim} and presented two studies: point source sensitivity estimates and simulations of $z\sim$\,2--3 emission-line star forming galaxies. Our main conclusions are as follows:

1. The E-ELT AO PSF is predicted to be a strong function of wavelength. We show the importance of incorporating this into our simulations. We use a novel parameterisation method to create specific PSFs at each wavelength of the data-cube, which then is convolved with its corresponding data-cube channel.

2. {\sc hsim} has been thoroughly crosschecked with existing implementations for other instruments, including the pipeline of \citet{Puech2010b} and the ESO SINFONI exposure time calculator, and these are consistent with our code.

3. We derive point source sensitivity estimates for observations with HARMONI using LTAO. The $20\,\rmn{mas}$ scale offers the greatest point source sensitivity in $H$-band, but the $10\,\rmn{mas}$ scale is more sensitive in $K$-band due to the increased thermal background. We also show that $H$-band observations are predominately read-out noise limited, whereas $K$-band is strongly thermal background limited.

4. We perform simulations of $z\sim$\,2--3 star forming emission-line galaxies and find that the E-ELT with HARMONI will be capable of obtaining velocity maps of these galaxies down to Milky-Way SFRs at these redshifts, with a factor $\sim5$ improvement in spatial resolution over current generation instruments. By deriving rotation curves we find that the E-ELT with HARMONI offers improved resolution data with a 10 fold improvement in observing efficiency compared to current telescopes, i.e. in equal observing times the E-ELT could observe 10 times as many objects. For equal observing times on the same galaxy the increased sensitivity of the E-ELT with HARMONI gives a factor of $\sim2$ improvement in dynamical mass estimates.

5. HARMONI with LTAO will provide exquisite resolved spectroscopy of $z\sim$\,2--3 galaxies, allowing the detection of individual star forming complexes and measurements of their properties. We demonstrate that it will be possible to detect star forming H{\sc ii} regions down to at least $\sim500\,\rmn{pc}$ radius in a galaxy of $\rmn{SFR}=4\,\rmn{M}_{\odot}\,\rmn{yr}^{-1}$ in a single night. We stress that this is only a representative example and a more detailed analysis of star forming clumps will be presented in a later work.

\section*{Acknowledgments}

We are grateful to the anonymous referee for helpful feedback which improved this paper. The authors also thank N. Schwartz for providing AO PSFs, M. Puech for helpful feedback while testing our pipeline, and J. Liske for help using {\it eltpsffit}. SZ, NT, SK, RH, MT and FC are supported by STFC-HARMONI grant ST/J002216/1. RH was also supported by STFC grant numbers ST/H002456/1 \& ST/K00106X/1. AMS acknowledges an STFC Advanced Fellowship (ST/H005234/1) and the Leverhulme foundation.

\bibliography{../MASTERBIB}
\bibliographystyle{mn2e_new}

\bsp

\label{lastpage}

\end{document}